\documentclass[%
 reprint,
 superscriptaddress,
 amsmath,amssymb,
 pra,
floatfix,
]{revtex4-2}

\usepackage{graphicx}
\usepackage{dcolumn}
\usepackage{bm}

\usepackage{natbib}
\usepackage{CJKutf8}
\usepackage{color}
\usepackage{ulem}
\usepackage{braket}

\begin{document}
\begin{CJK*}{UTF8}{gbsn}

\preprint{APS/123-QED}

\title{
{Field-Deployable Pressure Standard Based on a Compact Dual-Cavity Refractometer}
}

\author{Zhong-Liang Nie (聂中梁)}%
    \affiliation{Hefei National Research Center for Physical Sciences at the Microscale, University of Science and Technology of China, Hefei 230026, China}
\author{Jin Wang (王进)}
    \email{jinwang@ustc.edu.cn}
    \affiliation{Hefei National Laboratory, University of Science and Technology of China, Hefei, 230088, China}
\author{Zi-Fan Zhao (赵子凡)}
    \affiliation{State Key Laboratory of Chemical Reaction Dynamics, Department of Chemical Physics, University of Science and Technology of China, Hefei, 230026, China}    
\author{Chang-Le Hu (胡常乐)}%
    \affiliation{Hefei National Research Center for Physical Sciences at the Microscale, University of Science and Technology of China, Hefei 230026, China}
\author{Shui-Ming Hu (胡水明)}
    \affiliation{Hefei National Research Center for Physical Sciences at the Microscale, University of Science and Technology of China, Hefei 230026, China}
    \affiliation{Hefei National Laboratory, University of Science and Technology of China, Hefei, 230088, China}
    \affiliation{State Key Laboratory of Chemical Reaction Dynamics, Department of Chemical Physics, University of Science and Technology of China, Hefei, 230026, China}    
\date{\today}


\begin{abstract}
The next-generation pressure standard is moving toward optical-based, field-deployable systems. However, most existing optical refractometry pressure standards rely on bulky ultra-low expansion (ULE) cavities and complex feedback locking, limiting their portability and on-site applicability. Here, we present a miniaturized, transportable optical pressure manometer based on a dual-channel Fabry–Perot cavity machined from a single block of common fused silica. By employing a differential measurement between an evacuated reference cavity and a gas-exposed measurement cavity, common-mode errors such as thermal expansion and pressure-induced deformation are largely canceled, enabling the use of low-cost fused silica to achieve performance comparable to ULE. Radio-frequency scanning with Lorentzian fitting replaces conventional feedback locking, simplifying the optical design and improving robustness. Calibrated against a piston manometer, the device demonstrates a measurement repeatability of 3.4~ppm and a total uncertainty of \(u = \sqrt{(10.6\times10^{-6}p)^2 + (5.4~\mathrm{mPa})^2}\). The system can resolve periodic pressure fluctuations originating from the piston manometer and exhibits superior response speed. With its small footprint, portability, and independence from external frequency references, this fused-silica dual-cavity manometer offers a practical route toward on-site, quantum-traceable pressure calibration.
\end{abstract}

\maketitle

\section{Introduction \label{sec:Int}}

Traditional gas pressure standards mainly rely on conventional physical standards such as mercury manometers and static expansion systems. However, these methods have significant limitations: mercury is highly toxic, and the physical devices are susceptible to mechanical vibration and chemical corrosion, leading to insufficient long-term stability and reliability. With the development of the International System of Units (SI) towards traceability to fundamental constants~\cite{ICP2019}, optical interferometry has gradually become a mainstream technology to replace traditional physical standards~\cite{May2004-intro,Pendrill2004-intro,Egan2011-intro}. This method establishes a direct link between gas pressure and its microscopic properties, offering higher reproducibility and broader applicability.

The gas refractive index manometer based on optical interferometry is a high-performance pressure measurement technique. Its core principle relates the macroscopic gas pressure to the microscopic polarizability of atoms or molecules. Using the equation of state and the Lorentz–Lorenz formula, pressure measurement is converted into a precise measurement of the gas refractive index~\cite{Buckingham1974, Jousten2017, Rourke2019RIGT}. The refractive index is experimentally related to the longitudinal mode frequency of an optical resonator, and accurate gas pressure values can be retrieved through high-precision laser frequency measurements. This method not only avoids the inherent drawbacks of physical standards but also achieves a quantized, high-accuracy traceability of pressure or temperature. At present, both the National Institute of Standards and Technology (NIST, USA) and the National Institute of Metrology (NIM, China) have developed gas refractive index pressure standards based on nitrogen, with measurement uncertainties of 8.8 ppm ($k=2$)~\cite{Egan2016-NIST} and 23 ppm ($k=2$)~\cite{YANG2021Vaccum}, respectively, which are comparable to the performance of traditional methods. The core of the NIST and NIM laboratory-grade pressure standard is a specially shaped optical cavity. To maximally suppress cavity deformation caused by temperature changes, the cavity is made of ultra-low expansion glass (ULE) and incorporates a multi-layer temperature control structure. Prior to the work presented in this paper, we also built two ULE-based pressure standards in our laboratory, achieving an absolute measurement uncertainty of 20~ppm ($k=1$)~\cite{Xu2020WLXB-RIGT, Liu_measurement_2022, Nie_vaccu}. However, these systems integrate relatively complex optical designs and electronic control modules, resulting in a bulky overall footprint (0.5~m$\times$0.5~m$\times$1~m), sensitivity to environmental vibrations, and difficulty in portable or on-site deployment.

To realize miniaturization and portable integration of optical-based standards, enabling high-precision metrology to move out of national laboratories and into the field and industry, NIST and MKS Inc. jointly developed a portable FLOC pressure standard~\cite{NIST-MKS2021,NIST-MKS2025}. By employing a ULE cavity, optimizing laser locking and temperature control, and suppressing key noise sources such as residual amplitude modulation, they extended the effective measurement range from $0.01$~Pa to $400$~kPa and achieved a reading resolution uncertainty below $0.02\%$. The Swedish research group led by Axner developed a transportable optical Pascal standard based on an Invar dual-cavity structure and the gas modulation refractometry (GAMOR) method~\cite{GAMOR2020,GAMOR2024,GAMORCOMPARE2024}. Through multiple comparisons among three European national metrology institutes (PTB, INRiM, LNE), they verified that the device maintains stable performance under inter-laboratory transportation and different environmental conditions, with a maximum deviation of less than $8$~ppm and an average deviation of $4$~ppm from a piston manometer. More recently, complementary optical pressure standard approaches based on multi-reflection homodyne interferometry have been demonstrated at INRIM~\cite{Mari2023-UINT}, achieving a relative uncertainty of $10$ ppm at $100$~kPa in a compact footprint.

The demand for miniaturized, transportable pressure standards is driven by a broad range of practical applications. Precision pressure measurement is essential to numerous industrial sectors, including semiconductor manufacturing, where chamber pressure directly affects deposition rates, etch profiles, and device yield~\cite{Jousten2017}; aerospace testing and altimetry, where barometric accuracy directly impacts flight safety; and leak detection in sealed systems. Currently, industrial pressure transducers must be periodically sent to national metrology institutes for recalibration against primary standards --- a costly and time-consuming process. A compact, quantum-traceable pressure standard that can be operated on-site would eliminate this logistical bottleneck, enabling in-situ calibration of working gauges without interrupting production~\cite{Nie_vaccu}. The European EMPIR project ``QuantumPascal'' (18SIB04) has explicitly identified the miniaturization of optical pressure standards as a key step toward replacing mercury manometers and making the pascal realizable in industry~\cite{GAMORCOMPARE2024}. More recently, NIST has demonstrated dispersion barometry --- a two-color refractometric technique that realizes the pascal directly traceable to the SI unit of temperature, with the explicit goal of enabling industry and academia to establish their own optical pressure scales without recourse to external calibration services~\cite{Yang2025Dispersion}. Nevertheless, a low-cost, simple, and robust optical pressure gauge is highly desirable.

In this work, we present a transportable, miniaturized absolute pressure measurement system that can directly serve as a pressure standard for high-precision pressure calibration. The device features good mobility and portability, adapts to various on-site conditions, and supports real-time, in situ high-precision pressure calibration without the need for an external reference. Common fused silica is used instead of expensive ULE material, achieving comparable device performance. After calibration, the measurement repeatability is confirmed to be within the uncertainty range of the piston manometer. Moreover, the system can identify periodic fluctuations inherent to the piston manometer itself and exhibits superior reading sensitivity and temporal response.


\section{Method \label{sec:exp} }

The principle of the optical pressure manometer is to measure the refractive index of a working gas in a fixed-length cavity to obtain the gas density, and then combine it with the temperature to determine the gas pressure. The measurement formula can be expressed as ~\cite{equa2024}:
\begin{eqnarray}
 \frac{p}{RT} &=& \frac{2(n-1) }{3 (A_\varepsilon +A_\mu)} +
    (n-1)^2 \left[\frac{4 B_\rho(T)}{9(A_\varepsilon + A_\mu)^2} \right. \nonumber \\
 && \left. -\frac{A_\varepsilon^2+ 4 B_\varepsilon + 6 A_\varepsilon A_\mu}{9(A_\varepsilon+ A_\mu)^3}\right] \label{eq:p} 
\end{eqnarray}
where $R = 8.3144626~\mathrm{J~mol^{-1}~K^{-1}}$ is the molar gas constant, $A_{\varepsilon}$ is the molar electric polarizability, $A_{\mu}$ is the molar magnetic polarizability, $B_T$ is the second virial coefficient, and $B_{\varepsilon}$ is the second dielectric virial coefficient.

For a Fabry-P\'{e}rot optical resonator, the resonance mode frequency changes by $\Delta \nu$ before and after gas filling. The relationship between the gas refractive index and pressure is given by:
\begin{eqnarray}
n - 1 &=& \frac{\nu_f - \nu_0}{\nu_f} + \frac{p}{\kappa}, \label{eq:n-1} \\
\frac{1}{\kappa} & =& \frac{1}{3K} +d + \cdots,   \label{eq:kappa}
\end{eqnarray}
where $\nu_0$ and $\nu_f$ are the resonance frequencies of the measurement cavity before and after gas filling, respectively; $\kappa$ is the effective pressure correction coefficient. Main contributions to $\kappa^{-1}$ include: cavity spacer deformation described by the bulk modulus $K$, mirror deformation ($d$), mirror bending under pressure, and Gouy phase in the presence of gas~\cite{James2026, Zakrisson2025}. Note that this parameter depends only on the cavity material and structure, not on the type of working gas. For fused silica, we have typically~\cite{TAKEI2021} $K \approx 74~\mathrm{GPa}$ and $d^{-1} \approx 100~\mathrm{GPa}$. 

Combining Eqs.~\eqref{eq:p} - \eqref{eq:kappa}, the relationship between the frequency change and gas pressure can be obtained:
\begin{equation}
\frac{\nu_f-\nu_0}{\nu_f} = \left( \frac{1}{\alpha T} - \frac{1}{\kappa} \right) p - \frac{\beta}{\alpha^2T^2}p^2,
 \label{eq:nu} 
\end{equation}
where 
\begin{eqnarray}
\alpha &=& \frac{2R}{3(A_\epsilon+A_{\mu})}, \\
\beta &=& \frac{4R^2}{9(A_\epsilon+A_{\mu})^3} \left[ B_T - \frac{A_\epsilon^2 + 4B_\varepsilon + 6A_\epsilon A_\mu}{2(A_\epsilon + A_\mu)} \right].
\label{eq:AB} 
\end{eqnarray}


We adopt a design of dual-cavity optical pressure gauge (DC-Opt), and a double-channel resonant cavity is machined from a single fused silica block. The two channels serve as the ``reference cavity'' and the ``measurement cavity'', respectively. The reference cavity has been evacuated to a high vacuum and sealed; its mode frequency is influenced only by the inherent thermal expansion and pressure-induced deformation of the material, thus serving as a frequency reference. The measurement cavity has a U-shaped open cross-section and directly contacts the gas sample under test. Both ends of the two cavities are equipped with high-reflectivity mirrors and are firmly connected to the cavity body by a precision bonding process. Because the two cavities are machined from the same fused silica glass block and have identical geometric structures and materials, they are expected to exhibit similar responses to external pressure fluctuations and temperature changes. The effective correction coefficient in Eq.~\eqref{eq:nu} is replaced by $\kappa_{\mathrm{DC}}$. Therefore, measuring the frequency difference between the two cavities greatly suppresses common-mode deformation errors (such as changes in the bulk modulus of the cavity material and mirror deformation under pressure), improving the long-term stability and repeatability of pressure measurements.

\begin{figure}
 \centering
   \includegraphics[width=3.3in]{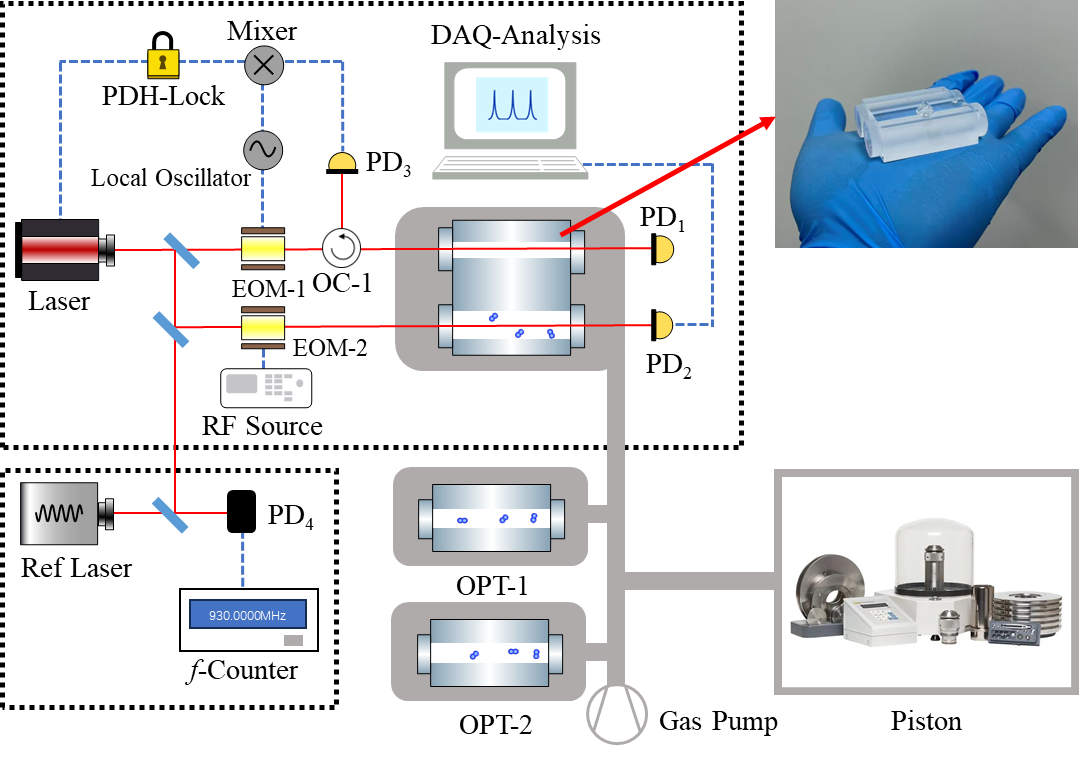}
\caption{Schematic diagram of the optical layout. The upper dashed-line box shows the configuration of the dual-channel optical pressure gauge (DC-Opt). The probe laser frequency is locked (through EOM-1) to the empty ``reference cavity'', and the resonance frequency of the ``measuring'' cavity is measured by scanning the sideband frequency driving EOM-2. The lower dashed-line box shows a reference laser system used to characterize the frequency drift of the DC-Opt. Other systems, including two individual optical gauges (Opt-1 and Opt-2) and a piston gauge, are also illustrated in the lower part. BS: beam splitter; EOM: electro-optic modulator; RF Source: radio-frequency source; PD: photodetector. Inset photographs show the compact fused-silica dual-cavity block (top right) and the laboratory piston manometer (bottom right).
}
\label{fig:optical-path}
\end{figure}

Figure~\ref{fig:optical-path} shows a schematic diagram of the experimental setup. The laser is a fiber laser (Precilasers, center frequency $\nu_0 \approx 191.358$~THz, linewidth $\delta\nu < 10$~kHz) whose tuning range exceeds the free spectral range of the optical cavity of about $3$~GHz. The laser output is divided into two spatial beams that enter the ``reference'' and ``measurement'' cavities, respectively. 
{For both channels, high-reflectivity mirrors with $R>99.9\%$ are used. The cavity length is $5$~cm, corresponding to a free spectral range of $3$~GHz. This yields a finesse of approximately $3000$ and a cavity mode linewidth of about $1$~MHz. The dual-channel optical cavity has overall dimensions of $5$~cm $\times$ $4$~cm $\times$ $1.8$~cm.} The main laser passes through an electro-optic modulator (EOM-1) and is coupled into the reference cavity, and the transmitted peak signal is observed by a photodetector PD1. The main laser frequency is locked to a reference cavity mode via a PDH-locking servo. The other laser path is frequency-shifted by a fiber EOM (EOM-2) and then coupled into the measurement cavity.
A RF source precisely controls EOM-2 so that its modulation sideband matches the frequency of the measurement cavity, and the RF frequency value is recorded in real time. A transmittance spectrum is obtained by scanning the sideband, and a Lorentzian fit of the spectrum yields the center frequency of the transmission peak. This value is the frequency difference $\nu_0$ between the measurement cavity and the reference cavity under vacuum. 
{The complete optical system (excluding the vacuum pump) weighs less than $10$~kg and occupies a volume of roughly $19$~cm $\times$ $16$~cm $\times$ $14$~cm, comparable in size to a shoebox.}

An external temperature control system of the optical cavity stabilizes the vacuum layer temperature at $24.4^\circ$C. 
{The outer surface of the temperature-control layer is wrapped with heating tapes and thermal insulation cotton. A TH10K thermistor is embedded on the inner side for in-loop temperature measurement, and a proportional-integral-derivative (PID) feedback loop is employed to maintain temperature stability. The temperature-control layer heats the ambient air outside the vacuum chamber, which in turn enables uniform heating of the vacuum chamber.} 
The internal cavity temperature is measured by a calibrated thermal sensor (Pt-100) attached to the cavity. 
{The temperature sensor is affixed to the outer surface of the vacuum layer with aluminum foil tape, and connected via a LEMO connector to an Anton Paar Millikelvin thermometer for real-time temperature readout.}

As the two cavities are machined from a single fused silica glass block, the cavity length changes and mirror deformations caused by temperature and pressure are largely canceled in the differential measurement. Therefore, $\nu-\nu_0$ mainly reflects the change in gas refractive index, and no active frequency stabilization of the reference cavity is required. Based on the change in frequency difference ($\nu-\nu_0$) before and after gas filling, combined with the real-time cavity internal temperature data, the corresponding pressure value is calculated according to Eq.~\eqref{eq:nu}. The center frequency of the resonance peak could be determined with an accuracy of a few kilohertz. This approach uses a single laser for both frequency reference and probing, replaces the active locking of the probe laser to the measuring cavity, simplifies the optical design, avoids the risk of losing lock during gas filling, and improves the system's immunity to disturbances.

\section{Results and Discussion \label{sec:res} }

\subsection{Calibration and uncertainty analysis}

\begin{figure}
 \centering
   \includegraphics[width=3.3in]{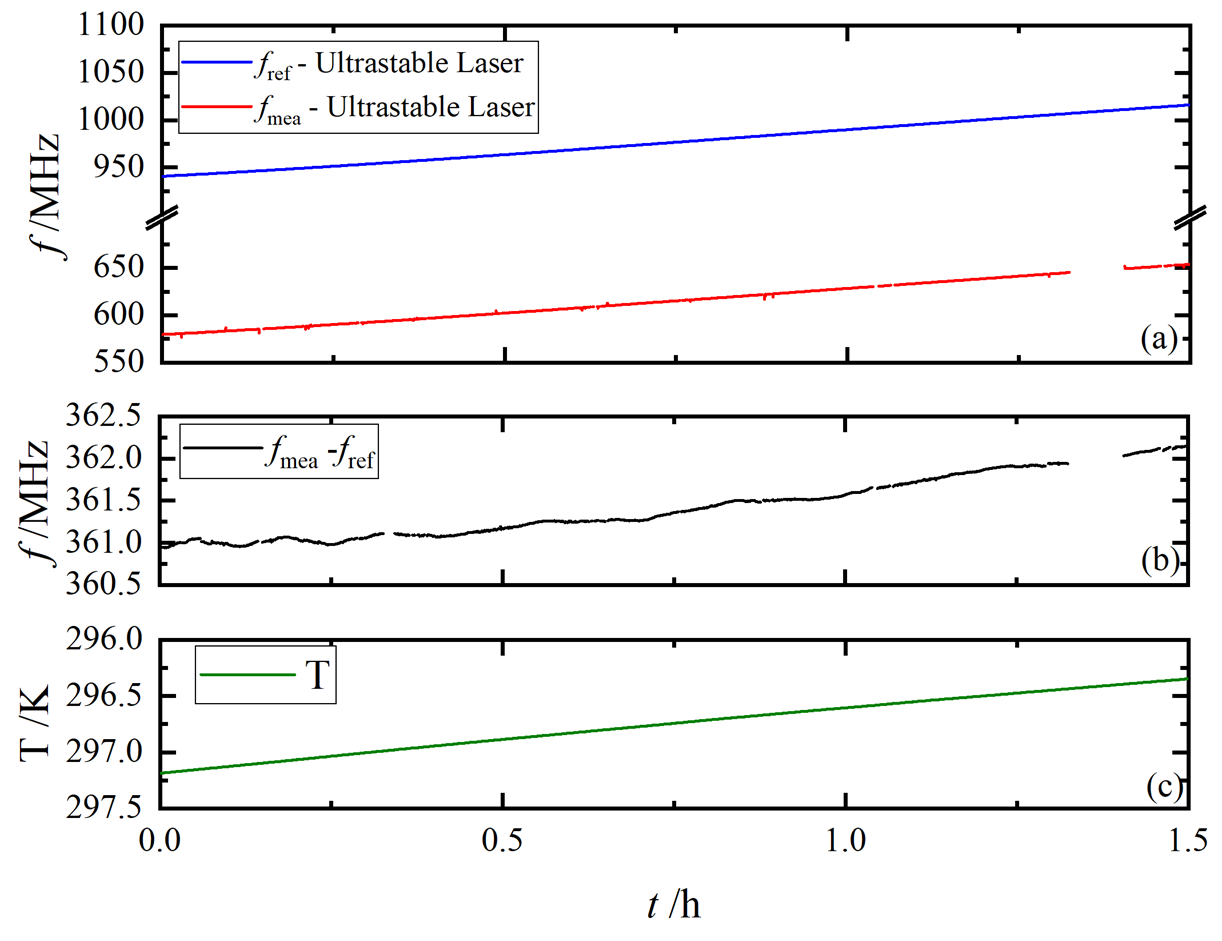}
\caption{Measurement of the dual-channel thermal expansion compensation effect.
}
\label{fig:thermal}
\end{figure}

For an optical pressure manometer, the key requirement is effective control of the optical cavity length variation, which mainly originates from the thermal expansion of the cavity material and directly affects the background level of the manometer. To suppress thermal expansion as much as possible, NIST uses ultra-low expansion (ULE) glass as the cavity material (coefficient of thermal expansion $\sim 3\times10^{-9}$~K$^{-1}$) and designs multi-layer temperature control structures to keep the cavity temperature fluctuation within a few mK. In this work, ordinary fused silica is used as the cavity material, which has a much larger thermal expansion coefficient (approximately $3\times10^{-7}$~K$^{-1}$). Owing to the dual-channel design, the thermal expansion effects of the measurement and reference cavities cancel each other in a common-mode manner, resulting in a reduced effective thermal expansion coefficient.
As shown in Fig.~\ref{fig:thermal}, we use a metrology-grade ULE ultra-stable reference laser as an external frequency reference, which has been examined to have a frequency drift of only $0.1$~Hz/s. The vacuum layer is always connected to a molecular pump to maintain high vacuum, and the frequency drifts of the reference and measurement cavities in the dual-channel fused silica glass cavity are monitored separately. We purposely change the temperature control over a period of $1.5$~h, and the platinum resistance thermometer (Pt-100) shows a temperature change of about $1$~K. The measured temperature-dependent frequency change rates of the reference and measurement cavities are $-99.76(4)$~MHz/K and $-97.75(5)$~MHz/K, respectively. The temperature-dependent frequency change rate of the difference between the reference and measurement cavities is only $-1.96(2)$~MHz/K. Thus, the dual-channel design suppresses the thermal expansion effect by about a factor of $50$, with an equivalent frequency drift coefficient of $1\times10^{-8}$~K$^{-1}$. If the overall cavity temperature drift can be controlled to $1$~mK, the self-referenced frequency drift will only be $2$~kHz, corresponding to a pressure change of only $4$~mPa for N$_2$, which meets the design requirements.

Another key factor affecting the measurement performance is the accurate evaluation of the cavity deformation caused by gas compression during repeated gas filling and venting. This directly affects the measurement repeatability during pressure measurements. The volumetric compression of the cavity material cannot be ignored, and the bulk modulus must be calibrated and its stability verified. To exclude this effect, we measured the frequency change of the sealed reference cavity. The DC-Opt chamber is repeatedly filled and vented, and the frequency change of the reference cavity is measured using the ultra-stable reference laser as an external frequency reference, from which the relationship between the deformation-induced frequency change and the external pressure is derived. The cavity temperature is monitored simultaneously. After subtracting the temperature drift contribution (about $30$~mK) using the previously obtained temperature-dependent frequency drift rate of the reference cavity, the frequency change solely due to the bulk modulus is obtained. A linear fit of the frequency change versus pressure yields a slope of $2.745(4)$~kHz/Pa. According to Eq.~\eqref{eq:kappa}, the slope value corresponds to an effective $\kappa$ value of $71.11$~GPa, agreeing with that estimated from the bulk modulus of fused silica glass and mirror deformation~\cite{TAKEI2021}.

\begin{figure}
 \centering
   \includegraphics[width=3.3in]{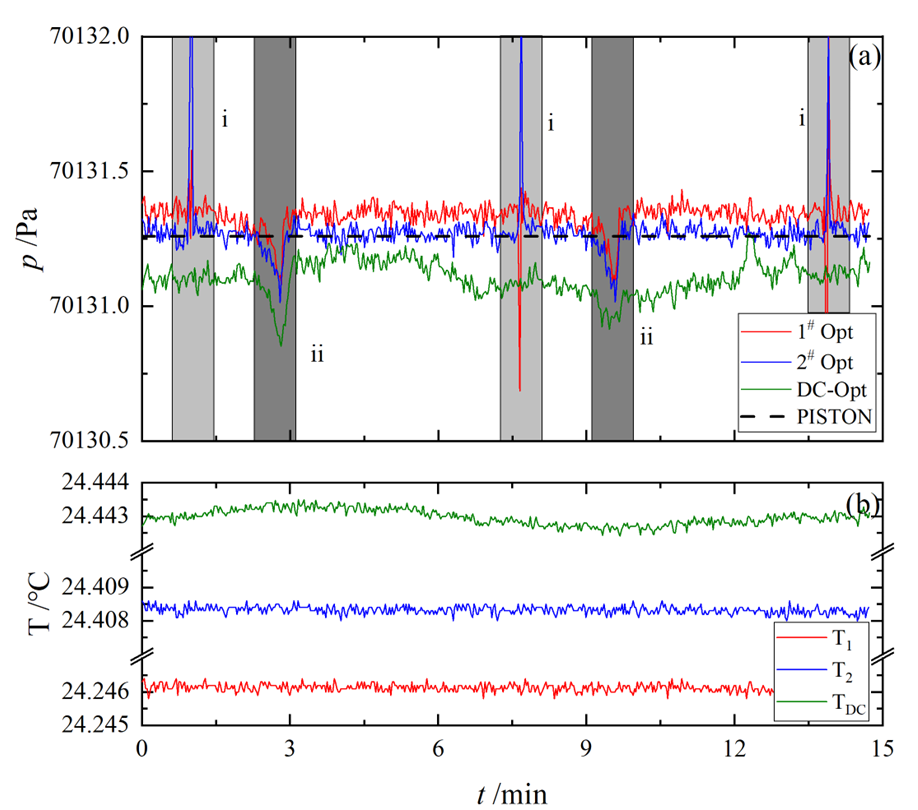}
\caption{(a) Simultaneous monitoring at a single pressure point; (b) Comparison of temperature readings. The shaded zones marked with ``i'' and ``ii'' were excluded in the data averaging, see the main text.
}
\label{fig:71kPa}
\end{figure}

As shown in Fig.~\ref{fig:optical-path}(a), a commercial piston manometer with a calibrated uncertainty of 6~ppm ($k=1$) is connected to the miniaturized optical pressure manometer through a high-purity gas line. The gas line also includes two laboratory-grade optical pressure manometers (Opt-1 and Opt-2) whose internal core cavities are made of single-channel ULE (detailed in Ref.~\cite{Nie_vaccu}). High-purity nitrogen (purity $99.9999\%$) is used as the working gas. At a pressure of $100~\mathrm{kPa}$, the piston manometer reading $p$ is compared with the frequency difference $\nu_f - \nu_0$ measured by the optical manometers. Using Eqs.~\eqref{eq:p} -- \eqref{eq:n-1}, the overall effect of the bulk modulus and mirror deformation on the refractive index measurement is calibrated. The effective coefficients for the two optical devices are determined as $\kappa_1 = 94.21~\mathrm{GPa}$ and $\kappa_2 = 92.44~\mathrm{GPa}$. 
For the dual-channel fused silica glass cavity, the bulk modulus contributions of the two cavities largely cancel, and the pressure-induced deformation coefficient mainly originates from the deformation of the reference cavity mirror. From the comparison with the piston manometer under $1~\mathrm{atm}$ of nitrogen, we obtain $\kappa_{\mathrm{DC}} = 99.285~\mathrm{GPa}$.

The measurement uncertainty of the interferometric optical pressure manometer has two categories. One includes factors linearly related to the measured pressure, such as the microscopic constants of the gas molecules, and is expressed as a relative value. The other one includes fixed biases in the measurement and is expressed as an absolute value (in mPa). Table~\ref{tab:uncertainty} shows the error budget for nitrogen as the working gas.

\begin{table}[htp]
\centering
\caption{Uncertainty budget for nitrogen as the working gas.} 
\begin{tabular}{cccc}
\hline
\hline
Paremeter& \multicolumn{2}{c}{Uncertainty} & Type\\
         & mPa\, \,   & $p\times 10^{-6}$ & \\
\hline
$A_\varepsilon$ &  & 7.3 & B \\
$A_\mu$         &  & 0.1 & B \\
$B_\varepsilon$ &  & 2.0 & B \\
$B_T$           &  & 2.8 & B \\
$\kappa$        &  & 6.0 & B \\
Gas impurity    &  & 1.0 & B \\
Compression hysteresis &  & 0.1 & B \\
Temperature     &  & 3.3 & A \\
Locking servo & 4 & & A \\
Outgassing/leakage& 3 & & A \\
Thermal Expansion & 2 & & A \\
\hline
sub-total, A & 5.4 & 3.3 & A\\
sub-total, B &     & 10.1 & B \\
Total  & \multicolumn{3}{c}{$u = \sqrt{(10.6\times 10^{-6}p)^2 +(5.4\,\mathrm{mPa})^2 }$} \\
\hline
\end{tabular} 
\label{tab:uncertainty}
\end{table}

At the operating wavelength of $1566~\mathrm{nm}$, the static polarizability is converted to the dynamic polarizability using the dipole oscillator strength distribution (DOSD) method~\cite{thakkar1992ab, bulanin1999dynamic}. For nitrogen, we have~\cite{Egan2016-NIST, Olney1997N2calculate, lesiuk_2024} $A_{\varepsilon} = 4.39573(3)~\mathrm{cm^3~mol^{-1}}$ and $A_{\mu} = -79.2(4)\times10^{-6}~\mathrm{cm^3~mol^{-1}}$.
The combined uncertainty of $A_{\varepsilon}+A_{\mu}$ contributes $7.3~\mathrm{ppm}$ (Type B) to the pressure measurement. According to the literature~\cite{Hohm1993, PFEganBrou}, we have $B_{\varepsilon} = 0.89(30)~\mathrm{cm^6~mol^{-2}}$ and $B_T = -5.01(9)~\mathrm{cm^6~mol^{-2}}$. 
From Eq.~\eqref{eq:nu}, within the range below 1~Bar, their contributions to pressure are about $2.0~\mathrm{ppm}$ and $2.8~\mathrm{ppm}$ (Type B), respectively. These parameter uncertainties depend on the working gas. 
For Ar or He, at the experimental wavelength, $A_{\varepsilon}^{\mathrm{Ar}} = 4.149543(24)~\mathrm{cm^3~mol^{-1}}$ and $A_{\varepsilon}^{\mathrm{He}} = 0.51774291(5)~\mathrm{cm^3~mol^{-1}}$~\cite{Gaiser2018PRL-A_epsilon, Olney1997N2calculate, puchalski2020Heqed}, corresponding to error contributions of $5.9~\mathrm{ppm}$ and $1.0~\mathrm{ppm}$, respectively.

The bulk modulus of the optical cavity has a linear relationship with pressure. In this experiment, it is calibrated using the piston manometer. The error contributed by this term to the pressure measurement is equivalent to the accuracy of the piston manometer used ($6~\mathrm{ppm}$, Type B).

The purity of argon used in the experiment is $99.9999\%$. Assuming $1~\mathrm{ppm}$ of atmospheric impurities, the resulting deviation is about $1~\mathrm{ppm}$ relative error (Type B).

The fused silica glass material exhibits a small compression hysteresis effect~\cite{Egan2016-NIST}, contributing about $0.1~\mathrm{ppm}$ relative error (Type B).

During the experiment, the cavity temperature is recorded in real time. The measurement uncertainty of temperature mainly comes from the reading accuracy of the Pt-100 platinum resistance thermometer and the temperature gradient between the thermometer placement and the actual cavity. The combined contribution of the temperature gradient deviation and the bulk modulus deviation has already been included in the previous term and is not repeated here. Therefore, only the repeatability needs to be considered. Figure~\ref{fig:71kPa} shows the temperature readings of three optical manometers in a typical measurement, indicating that the temperature fluctuation level of the DC-Opt device is about $1~\mathrm{mK}$. The contribution of this temperature term to the pressure measurement error is $3.3~\mathrm{ppm}$ (Type A).

The laser frequency locking system has a systematic drift. In the experiment, the drift range is about $2~\mathrm{kHz}$, which translates to a pressure measurement error of about $4~\mathrm{mPa}$ (Type A).

After evacuating the cavity and closing the metal valve, the frequency drift caused by outgassing and leakage is measured. The maximum deviation introduced to the pressure measurement is $3~\mathrm{mPa}$ (Type A).

The measured effective thermal expansion coefficient of the optical cavity is $1.02(4)\times10^{-8}~\mathrm{K^{-1}}$. Considering the system temperature fluctuation of about $1~\mathrm{mK}$, the corresponding laser frequency change after correction is about $1~\mathrm{kHz}$, which corresponds to a pressure measurement influence of about $2~\mathrm{mPa}$ (Type A).

Combining all contributions, the total uncertainty of the dual-cavity optical pressure manometer is $u = \sqrt{(10.6\times10^{-6}p)^2 + (5.4~\mathrm{mPa})^2}$, in which the statistical uncertainty (Type-A) is only 0.34~Pa at 1~Bar.

\subsection{Comparison measurement of the optical pressure manometer}

\begin{figure}
 \centering
   \includegraphics[width=3.3in]{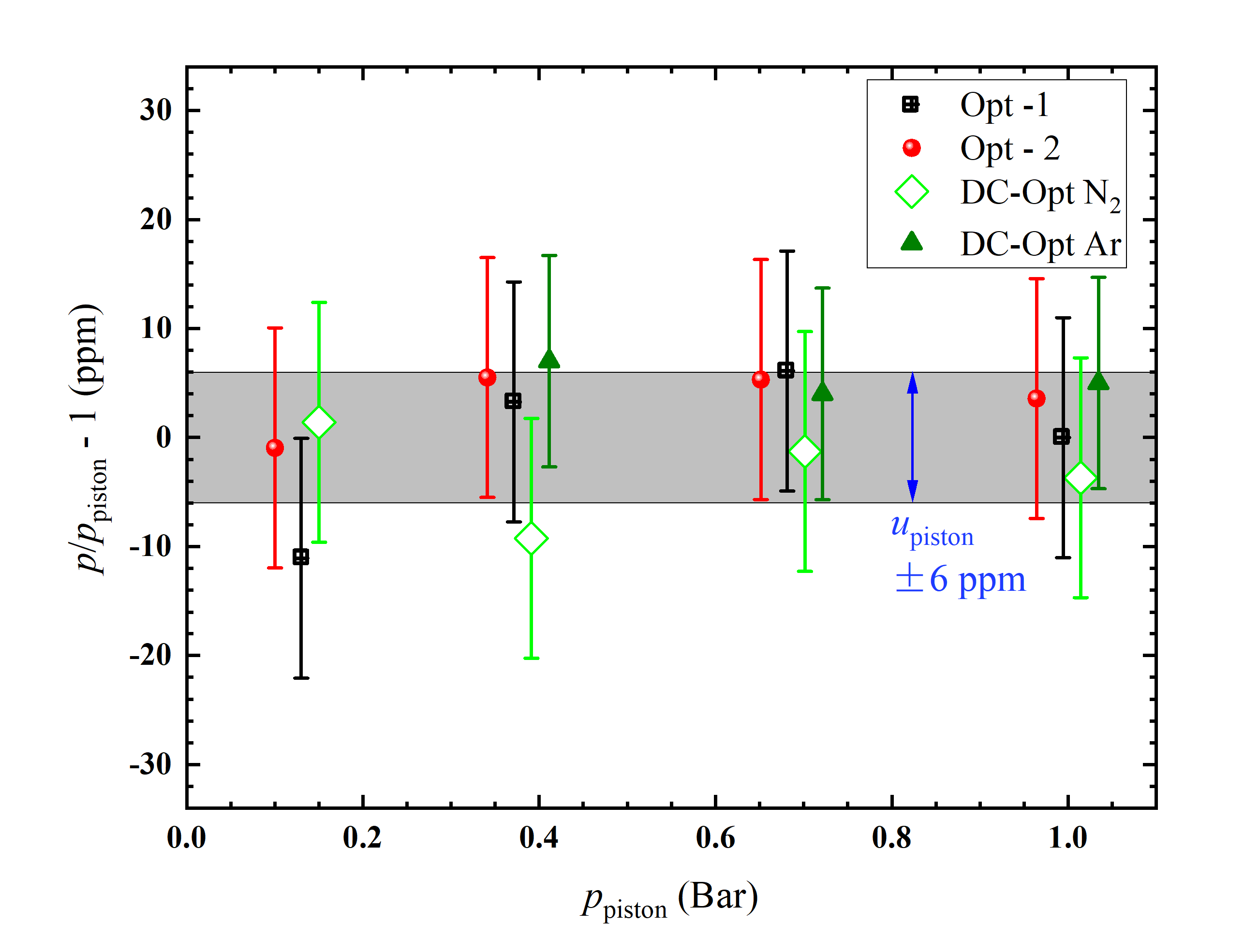}
\caption{Comparing pressure readings from optical (Opt-1, Opt-2 and DC-Opt) manometers with the piston values, using N$_2$ and Ar as working gases.
}
\label{fig:comparison}
\end{figure}

To verify the measurement uncertainty of the calibrated optical pressure manometers, high-purity nitrogen is again introduced to a pressure of $71~\mathrm{kPa}$, and after equilibration the readings are recorded. As shown in Fig.~\ref{fig:71kPa}(a), the deviations of the optical systems Opt-1 and Opt-2 from the piston manometer are about $0.1~\mathrm{Pa}$. The deviation of the dual-channel optical pressure manometer DC-Opt from the piston reading is about $-0.2~\mathrm{Pa}$ ($3~\mathrm{ppm}$), with a slow drift of amplitude $0.1~\mathrm{Pa}$ and a period of about ten minutes. The DC-Opt system, designed for simplicity and compactness, has only a single-layer temperature control, using air between the control layer and the vacuum chamber for temperature regulation, making it more susceptible to ambient temperature fluctuations. In contrast, Opt-1 and Opt-2 have more sophisticated multi-layer temperature control designs, including an active temperature control layer outside the vacuum chamber and a thermal insulation layer, so their temperature stability is significantly better than that of DC-Opt. Figure~\ref{fig:71kPa}(b) shows that the temperature reading noise of Opt-1 and Opt-2 is about $0.5~\mathrm{mK}$, and the long-term drift over several hours is less than $2~\mathrm{mK}$, while DC-Opt temperature readings exhibits fluctuations of about $1~\mathrm{mK}$. It is noted that this temperature drift is clearly correlated with the pressure reading fluctuation.

In the monitoring shown in Fig.~\ref{fig:71kPa}(a), three optical manometers exhibit two types of periodic transient fluctuations. The first type is a spike-like disturbance (marked ``i'' in the figure) lasting only a few seconds. The second type (marked ``ii'') lasts about one minute with an amplitude of $0.2~\mathrm{Pa}$. By synchronously monitoring the piston operation, we attribute these fluctuations to the periodic operating states of the piston manometer. The piston manometer operates on the principle of hydrostatic equilibrium: standard weights are loaded onto a precision piston; when the upward force generated by the gas pressure balances the total gravitational force of the piston and weights, the piston floats and rotates at a constant speed~\cite{Ehrlich1994PISTON, Durgut_2018piston}. The spike-like disturbance occurs when the gas replenishment system of the piston manometer briefly supplies gas to the volume below the piston, causing a temporary pressure imbalance. The second type, lasting about one minute, likely corresponds to the re-acceleration of the piston ring after a rotation stall.

Notably, the response of the optical manometers to these fluctuations differs due to their different operating principles. For the short-lived spike (type i), Opt-1 and Opt-2, which employ the Pound-Drever-Hall (PDH) frequency locking scheme, temporarily lose lock and produce distorted readings. In contrast, DC-Opt, which determines the resonance frequency by scanning the transmission peak, has a slower response and cannot follow such rapid pressure changes. For the longer-duration fluctuation (type ii, $\sim 1$~min), all three optical manometers have sufficient response speed and show consistent trends. Since this work focuses on steady-state measurement accuracy of the optical pressure manometer, we exclude data points from these transient periods (shaded areas in the figure) in subsequent comparisons between the optical and piston manometers. Nonetheless, the results demonstrate that optical manometers (especially those with PDH locking) can clearly detect rapid tiny pressure fluctuations and offer faster response capability than the piston manometer.

To verify the repeatability of the optical pressure manometer over its measurement range, Opt-1, Opt-2, and DC-Opt are compared with the piston manometer at several pressure points from $10$ to $100~\mathrm{kPa}$. The piston manometer reading is taken as the reference to determine the deviation of the three optical manometers. The results are shown in Fig.~\ref{fig:comparison}. Within the test range, the relative deviations of the three optical manometers are all comparable to the nominal measurement uncertainty of the piston manometer. To verify the measurement capability of the DC-Opt system, argon is also used as the working gas for repeated measurements. The results are also shown in Fig.~\ref{fig:comparison}, with deviations at the same level. It is worth noting that when using argon, the effective pressure correction parameter $\kappa$ is still the value calibrated under $1~\mathrm{atm}$ of nitrogen. Thus, within the experimental error, this correction is indeed independent of the gas type.

It is worth mentioning that the GAMOR system of the Swedish group uses periodic gas modulation to mitigate drift and suppress thermodynamic disturbances~\cite{Rubin2022Metrologia}. The complementary design presented here instead relies on the intrinsic common-mode rejection of the dual-channel fused-silica block and on RF scanning of the cavity mode, which simplifies the hardware and keeps the time response within a few seconds.



\section{Conclusion \label{sec:con}}

In this work, we have designed and built a miniaturized optical pressure manometer based on a dual-channel fused-silica resonant cavity. By using a common-block dual-cavity design and differential measurement, common-mode errors such as thermal expansion and pressure-induced deformation are effectively suppressed, allowing ordinary fused silica to achieve a temperature stability comparable to that of ultra-low expansion glass. The RF scanning and Lorentzian fitting method for extracting cavity mode frequencies avoids complex frequency-locking circuits and improves the system's immunity to disturbances. After calibration, the pressure manometer demonstrates a drift below 0.3~Pa, a total uncertainty of 1.06~Pa at 100~kPa, and a response time of a few seconds. Comparison experiments with a piston manometer confirm good agreement over the tested pressure range.

Compared with the FLOC device developed by NIST~\cite{NIST-MKS2025} and the GAMOR system by the Swedish research group~\cite{GAMORCOMPARE2024}, our dual-cavity miniaturized optical pressure manometer achieves comparable measurement accuracy under both N\(_2\) and Ar working gases. The accuracy is mainly limited by the piston manometer used for calibration and by temperature drift. With further improvements in temperature control, we can expect enhanced accuracy. Owing to the use of ordinary fused silica as the substrate and optical contact bonding for mirror attachment, our design significantly reduces cost and holds the potential for operation over a wide temperature range. These features are of great significance for realizing pressure metrology in real-world, on-site applications. The device is portable, movable, and does not require an external reference, making it a practical candidate for a pressure standard in field environments.

\begin{acknowledgments}
This work was jointly supported by 
 the Independent Deployment Project of HFNL (Grant No. ZB2025010500),
 and the Chinese Academy of Sciences (Grant Nos. XDA0520304 and XDB0970100).  

\end{acknowledgments}

\end{CJK*}

\bibliographystyle{iopart-num}
\bibliography{RIGT-H2}
\end{document}